\documentclass[%
 reprint,
 amsmath,amssymb,
 aps,
]{revtex4-2}
\usepackage{mathrsfs}
\usepackage{graphicx}
\usepackage{dcolumn}
\usepackage{bm}
\usepackage[none]{hyphenat}
\usepackage{epstopdf}
\usepackage{epstopdf}

\begin{document}

\emergencystretch 3em

\preprint{APS/123-QED}

\title{Role of tensor forces in nuclei}

\author{Yu.P.Lyakhno}

\affiliation{%
 National Science Center "Kharkiv Institute of
Physics and Technology" \\ 61108, Kharkiv, Ukraine\\}%

\date{\today}

\begin{abstract}
Recently, ground-state calculations for the lightest nuclei have
been performed using high-precision data on realistic internucleon
forces. In the present work, these results are employed to describe
certain properties of nuclei with mass number $A>4$. Accounting for
tensor forces leads to the conclusion that four subsystems within
the nucleus with zero nucleon orbital angular momenta primarily
combine into a $^1S_0$ cluster. Subsystems with non-zero orbital
angular momenta also form clusters with lower potential energy. Such
an approach provides a consistent explanation for the $^8$Be
lifetime, the Hoyle state, the sequential mechanism of reactions
with $\alpha$-particle emission, the reaction threshold shift, and
other phenomena. While the assumption of a one-dimensional effective
nucleon interaction implies that the nucleus contains a "force
center" relative to which nucleons acquire orbital angular momenta.
Our approach does not predict the existence of such a "force center"
in the nucleus.

\begin{description}

\item[PACS numbers]
21.30.Fe, 21.90.+f

\end{description}
\end{abstract}

\pacs{Valid PACS appear here}

\maketitle

\section{Introduction}

Currently, high-precision data are available on the realistic
two-nucleon forces acting within the nucleus. For illustration,
Fig.~1 shows the data for the CD-Bonn potential \cite{1}. The
meson-exchange model of nuclear forces was employed to determine the
shape of these curves. Experimental data on the deuteron binding
energy were also utilized. These curves were fitted using the least
squares method to experimental data on various observables of
elastic \(pp\) and \(np\) scattering in the energy range up to \(E_N
= 350\)~MeV. The partial-wave expansion of the total angular
momentum for the two-nucleon system was truncated at \(J = 4\)
(notation: \(^{2S+1}L_J\), where \(S\) is the spin and \(L\) is the
orbital angular momentum of the two-nucleon system). The \(\chi
^{2}\) value was approximately 1 per datum. Fig.~1 shows the \(np\)
scattering phase shifts for (a) \(T = 0\) and (b) \(T = 1\). The
positive phase value corresponds to attraction between the nucleons,
whereas the negative phase value implies repulsive interaction
between the nucleons. It can be seen from Fig.~1 that the most
intense interaction between nucleons occurs in states with zero
orbital angular momenta.

The possibility of nucleon modification within the nucleus is
discussed in the literature \cite{2}. It is hoped that if such
modification exists, it may also occur during elastic two-nucleon
scattering and, consequently, can be accounted for
phenomenologically.

Three-nucleon forces 3{\it NF's} also act within the nucleus, with
their contribution to the binding energy being approximately $\sim
10\%$ of that from the {\it NN} interaction. This issue lies beyond
the scope of the present paper.

\begin{figure}[!h]
  \centering
  \includegraphics[width=90mm]{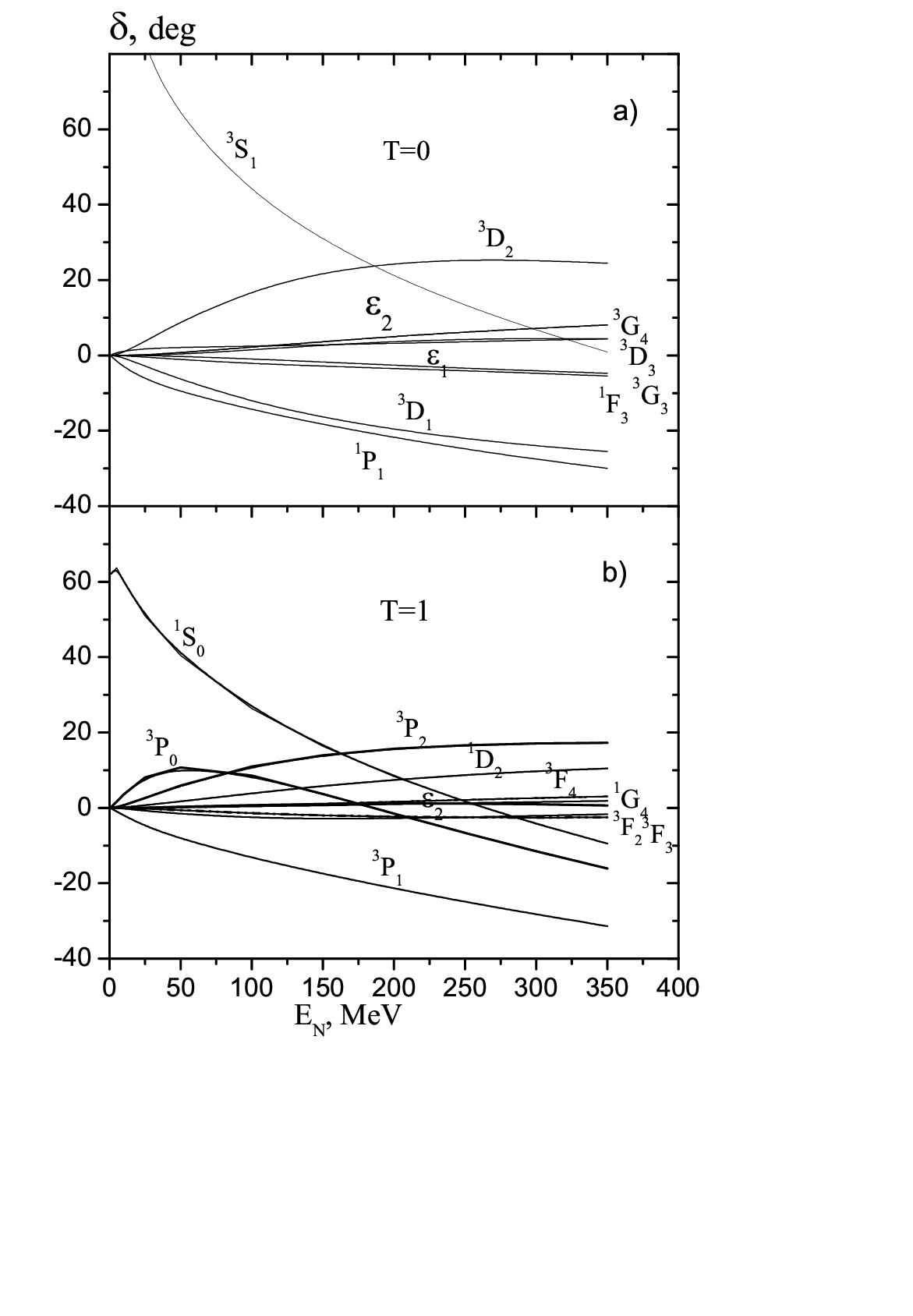}
  \vspace{-80pt}
  \caption{Phases $\delta$ and mixing coefficients $\varepsilon$ for
  {\it np} scattering with a) T=0 and b) T=1 for {\it NN} potential
  CD-Bonn up to the total angular momentum of the two-nucleon system
  J$\leq$4. }
\end{figure}

In Ref.~\cite{3}, the binding energies and probabilities of states
with non-zero nucleon orbital angular momenta were calculated for
the ground states of the lightest nuclei. The calculations employed
the CD-Bonn \cite{1} and Argonne AV18 \cite{4} two-nucleon (NN)
potentials, as well as the Tucson-Melbourne TM \cite{5} and Urbana
IX \cite{6} three-nucleon 3{\it N} potentials. The computations were
performed using the Faddeev \cite{7} and Yakubovsky \cite{8}
methods. A powerful computer was used for these calculations. The
results are presented in Table~I, adapted from Ref. \cite{3} (the
table is abbreviated).

%\end{multicols}
%\vspace{10mm}
%\begin{minipage}{165 mm}
\begin{center}
\vspace{3 mm} \emph{\textbf{Table 1.}} {\it S}, $S^{\prime}$, {\it
P} and {\it D} state probabilities for $^4$He and $^3$He nuclei.

%\hspace{3 mm}
{\small \begin{tabular}[t]{c |c c c c|c c c c} \hline\hline
& \multicolumn{4}{c|}{$^4$He} & \multicolumn{4}{c}{$^3$He}  \\
%\cline{2-9}
Model &  $S\%$ &  $S^{\prime}\%$ & $ P\%$ & $ D\%$  & $S\%$ &  $S^{\prime}\%$ & $ P\%$ & $ D\%$ \\
\hline
 AV18 & 85.45 & 0.44 & 0.36 & 13.74 & 89.95 & 1.52 & 0.06 & 8.46 \\

 CD-Bonn & 88.54 & 0.50 & 0.23 & 10.73 & 91.45 & 1.53 & 0.05 & 6.98 \\

 AV18+UIX & 82.93 & 0.28 & 0.75 & 16.04 & 89.39 & 1.23 & 0.13 & 9.25 \\

 CD-Bonn+TM & 89.23 & 0.43 & 0.45 & 9.89 & 91.57 & 1.40 & 0.10 & 6.93 \\
\hline\hline
\end{tabular}}
\end{center}

As can be seen from Table~I, the \(^3\text{He}\) (\(^4\text{He}\))
nucleus spends approximately $\sim$90\% ($\sim$85\%) of the time in
states with zero nucleon orbital angular momenta. For the remainder
of the time, these nuclei are in various configurations with
non-zero orbital angular momenta.

\section{Discussion of the structure of nuclei with mass number $A>4$ based
on realistic {\it NN} interaction}

Physicists often state that the interaction of nucleons within a
nucleus differs from that of free nucleons \cite{9} and, for
computational convenience, introduce simpler one-dimensional
potentials. Such a simplification of realistic internucleon forces
may lead to questionable conclusions regarding nuclear structure. In
the present work, we have utilized the results of precise
calculations of the properties of nuclei with mass number \(A \leq
4\), performed on the basis of realistic internucleon forces, to
investigate the properties of nuclei with \(A > 4\).

As can be seen from Fig. 1, the realistic interaction between
nucleons depends on four fundamental variables. This can be
formulated as follows: the interaction between nucleons unfolds
within a four-dimensional generalized configuration space. The first
axis of this space represents the inter-nucleon distance, where the
coordinate takes a continuous range of values from \(0\) to \(\infty
\). The remaining discrete axes govern the quantum states of the
system: the second axis is the orbital angular momentum \(L\) (\(0,
1, 2...\)), the third axis is the total spin \(S\) (\(0\) or \(1\)),
and the fourth axis is the isospin \(T\) (\(0\) or \(1\)).
Consequently, nuclear interaction cannot be adequately described
within purely spatial or spacetime coordinates. Upon the collision
or approximation of nucleons, a novel internal configuration space
with its own distinct axes of anisotropy is physically manifested.
It is precisely along these internal axes that the real tensor
forces emerge. Accordingly, the nuclear interaction of nucleons must
be formulated as a non-central tensor interaction operating across
these multiple dimensions. It is fundamentally impossible to
substitute this multi-component tensor interaction with a
conventional simplified vector or central potential.

 The presence of tensor forces does not alter the external
characteristics of the nucleus, such as the total energy \(E\),
total angular momentum \textbf{\textit{J}}, and nuclear parity.
Tensor forces modify the individual orbital angular momenta  of the
nucleons $\textbf{\textit{$l_i$}}$, which leads to a change in the
total orbital angular momentum of the nucleus  $\textbf{\textit{L}}
= \sum_{i=1}^{A-1}\textbf{\textit{l}}_i$, and to changes in the
nucleons potential and kinetic energies and the mass defect. The
balance between these quantities is established such that the
nuclear binding energy is conserved. The total orbital angular
momentum \textbf{\textit{L}} can take values that satisfy the
conservation law for the total angular momentum of the nucleus,
\textbf{\textit{J = L}}+\textbf{\textit{S}}. In the general case,
the total nuclear spin
 $\textbf{\textit{S}} = \sum_{i=1}^A
\textbf{\textit{s}}_i$ can take values \(0 \leq S \leq A/2\) if the
mass number \(A\) is even, and \(1/2 \leq S \leq A/2\) if \(A\) is
odd.

The tensor forces provide their smallest contribution, approximately
5\%, to the structure of the deuteron, in which only two states {\it
S} and {\it D} with parallel nucleon spins are allowed. Each
additional nucleon in the nucleus increases the probability of the
tensor interaction contribution. Consequently, medium-mass and heavy
nuclei may spend a significant portion of the time in configurations
determined by tensor forces.

All nucleon states in the nucleus must satisfy the Pauli exclusion
principle, which can be formulated as follows: any point in the
four-dimensional space of a system of half-integer spin particles
can be occupied by no more than one particle. This law precludes the
formation of two or more identical nucleon subsystems within the
nucleus. Ignoring the Pauli exclusion principle may lead to
erroneous interpretations of nuclear structure and the mechanisms of
nuclear reactions \cite{10}.

It can be seen from Fig.~1 that nucleon subsystems with zero orbital
angular momenta possess the highest potential energy (in absolute
value). Four such nucleon subsystems can exist within a nucleus: two
\(pp\) and \(nn\) interactions with isospin \(T = 1\), and two
\(pn\) interactions with \(T = 1\) and \(T = 0\). The binding energy
of a three-nucleon system is approximately 2.6~MeV/nucleon, while
for a four-nucleon system, it is about 7~MeV/nucleon. Therefore, the
four possible subsystems with zero orbital angular momenta primarily
combine into a \(^{1}S_{0}\) cluster (Fig.~2a). Intuitively, the
\(^{1}S_{0}\) cluster resembles the "S-shell" predicted by the
nuclear shell model (NSM). However, a fundamental difference between
the \(^{1}S_{0}\) cluster and the "S-shell" is that the former does
not require a "force center" for its formation. It can move freely
throughout the nuclear volume, exchange nucleons, decay into several
parts according to Young tableaux, and even vanish entirely.

The NSM does not predict the existence of states with non-zero
nucleon orbital angular momenta in the \(^{4}\text{He}\) nucleus, as
listed in Table~1. The NSM predicts states only with zero nucleon
orbital angular momenta, asserting that all other states are
forbidden according to the Pauli exclusion principle. At the same
time, the NSM does not dispute the fact that a \(D\)-state exists in
the deuteron. This situation can be explained by the fact that the
assumption of a one-dimensional effective nucleon interaction within
the nucleus is a very rough approximation of realistic internucleon
forces. As a result, we did not find a "force center" in the
nucleus. Physicists have previously expressed doubts regarding the
existence of a "force center" in the nucleus \cite{11}.

\begin{figure}[h]
\noindent \centering {
\includegraphics[width=90mm]{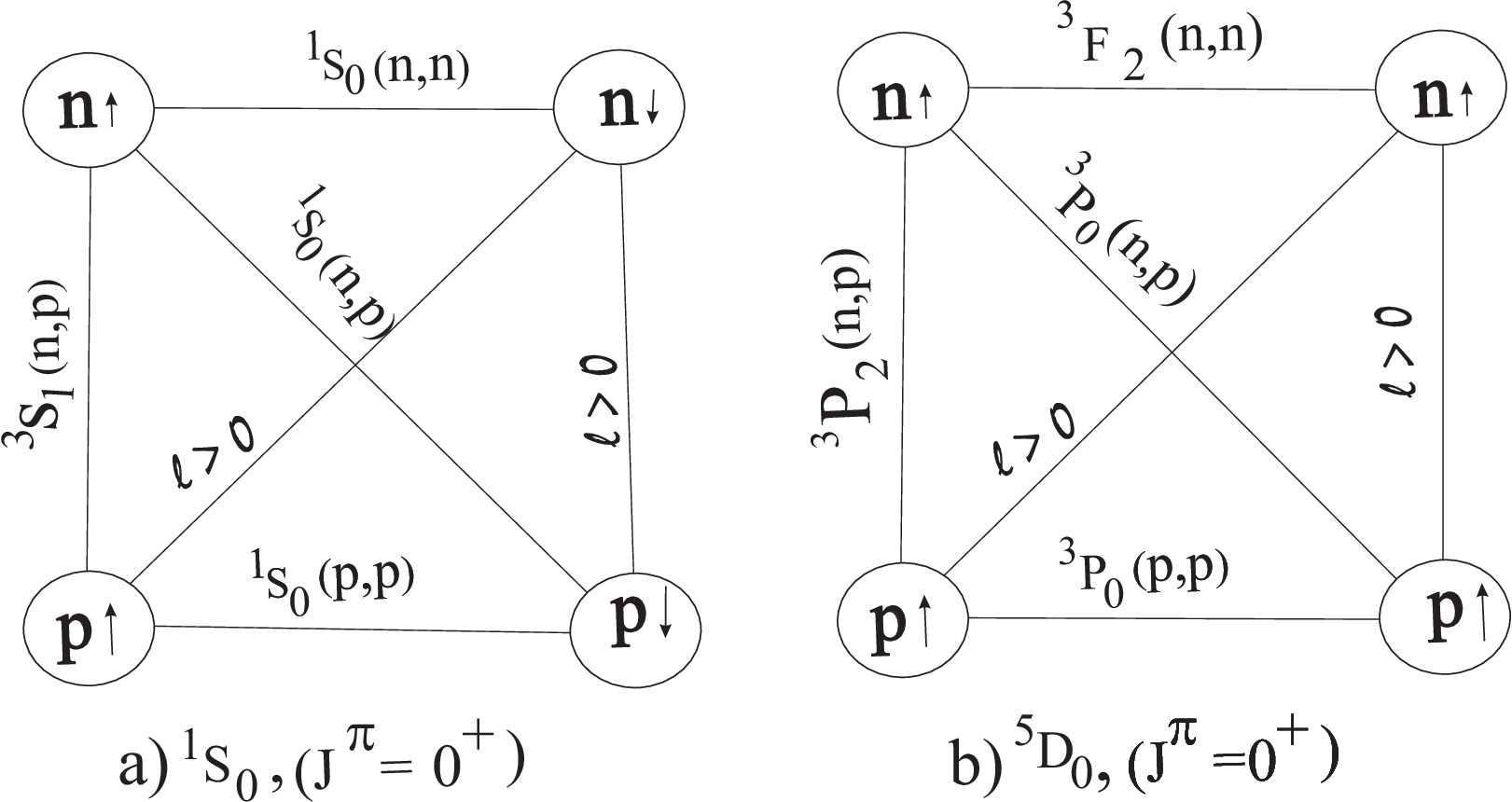}
}\vspace{50pt} \caption{Possible configurations of nucleons in a) S
and b) D clusters in nuclei. Arrows indicate the direction of
nucleon spins.} \label{figCurves}
\end{figure}

$^1S^{\prime}_0$ clusters consisting of four nucleons with non-zero
orbital angular momenta and a total cluster orbital momentum of {\it
L}=0 may also arise in the nucleus. However, according to the
calculation \cite{3}, the clusters with the next highest potential
energy after the $^1S_0$ cluster may be the $^5D_0$ clusters
(Fig.~2b). These clusters are formed from subsystems with non-zero
nucleon orbital angular momenta; consequently, several such clusters
may exist within the nucleus.

It is known that the $^8$Be nucleus decays into two
$\alpha$-particles, and the mass of the $^8$Be nucleus exceeds the
mass of two $\alpha$-particles: M($^8$Be)-2M($\alpha$)=0.092~MeV.
The decay width of the $^8$Be nucleus is $\Gamma$=5.57~eV. This
corresponds to a $^8$Be lifetime of
$\tau$($^8$Be)$>$10$^6$$\tau_{nucl}$, where $\tau_{nucl}$$\sim 10^{
-22}$ s is the characteristic nuclear time scale. This leads to a
paradox: a system of particles with a positive binding energy
possesses an exceptionally long lifetime by nuclear standards.

This paradox can be explained as follows. First, it is necessary to
account for the fact that, according to the Pauli exclusion
principle, the $^8$Be nucleus cannot consist of two
$\alpha$-particles. Furthermore, one may assume that the $^8$Be
nucleus consists of {\it S}- and $\it D$-clusters. The potential
energy of the {\it D}-cluster is lower than that of the {\it
S}-cluster; consequently, it is more massive than an
$\alpha$-particle. This result may lead to a negative binding energy
for the $^8$Be nucleus: M($^8$Be)-[M(S)+M(D)]$<$0. In a certain
nucleon configuration, an {\it S}-cluster is emitted. Once the {\it
S}-cluster moves several fm away from the parent nucleus, it
transforms into a regular \(\alpha \)-particle. The remaining {\it
D}-cluster then transforms into an {\it S}-cluster and subsequently
into a regular $\alpha$-particle. It should be noted that during
these transformations, only the internal states of the clusters
change.

Currently, it is assumed by default that the threshold for the decay
of the $^{12}$C nucleus into three $\alpha$-particles is at a photon
energy of $E_{\gamma}$=3M($\alpha$)-M($^{12}$C)=7.27~MeV. The total
angular momentum and parity of this level are $J^\pi$=$0^+$.
However, based on experimental data obtained, for instance, from the
study of the $^{14}$N($d,\alpha$)~$^{12}$C$^*$ reaction \cite{12},
it was concluded that transitions from this level to the ground
state of the $^{12}$C nucleus are not observed. Instead, transitions
are observed from a level with an energy of
$E_{\gamma}$=7.68$\pm$0.03~MeV. Hoyle calculated this level based on
astrophysical observations \cite{13}. Physicists denote this level
as $J^\pi$=$0^+_2$.

According to the Pauli exclusion principle, the $^{12}$C nucleus
cannot consist of three \(\alpha \)-particles. If the $^{12}$C
nucleus were composed of three $\alpha$-particles, its
disintegration threshold would indeed be at a photon energy of
$E_{\gamma}$=7.27~MeV. We hypothesized that transitions from the
$E_{\gamma}$=7.27~MeV level of the $^{12}$C nucleus are not observed
because this level does not exist. Given that a {\it D}-cluster is
heavier than an $\alpha$-particle, the threshold for the breakup of
the $^{12}$C nucleus into three $\alpha$-particles must be higher
than $E_{\gamma}$=7.27~MeV. According to experimental data, this
level may be at a photon energy of $E_{\gamma}$=7.65~MeV. Using the
relation $E_{\gamma}$=M($\alpha$)+2[M($\alpha$)+$\Delta$M({\it
D})]-M($^{12}$C)=7.65~MeV, we estimated how much smaller the mass
defect of a {\it D}-cluster is compared to that of an
$\alpha$-particle: $\Delta$M({\it D})=(7.65-7.27)/2=0.19~MeV. Since
\(D\)-clusters possess different potential energies, the calculated
difference in mass defects represents an average value for the two
{\it D}-clusters in the $^{12}$C nucleus. Nevertheless, this result
allows us to estimate the binding energy of the $^8$Be nucleus:
$E_{bind}$($^8$Be)=0.092-0.19=-0.098~MeV, and, accordingly, to
explain the long lifetime of this nucleus.

The disintegration of the $^{12}$C nucleus into three
$\alpha$-particles follows the scheme described above. After the
{\it S}-cluster moves to infinity, it becomes a regular
$\alpha$-particle. Subsequently, one of the {\it D}-clusters
transforms into an {\it S}-cluster and is also emitted. It should be
noted that during the flight to the detectors, each of these
$\alpha$-particles may transform into a {\it D}-cluster several
times \cite{3}. This sequential reaction mechanism is consistent
with numerous experimental data \cite{14}.

The mechanism of the $^{12}$C photodisintegration into three
$\alpha$-particles differs significantly from the mechanism of {\it
S}- or {\it D}-cluster emission from the $^8$Be nucleus. For
instance, the energy of the first $\alpha$-particle can exceed that
of the other $\alpha$-particles by an amount depending on the
reaction threshold shift \(\Delta E_\gamma = 0.38\)~MeV.

Based on the obtained data, we also estimated the threshold for the
\(^{16}\text{O}\) photodisintegration into \(4\alpha\)-particles as
follows: $E_{\gamma}$=4M($\alpha$)-M($^{16}$O)=14.44~MeV. To this
value, we must add 3$\Delta$M({\it D}). As a result, we obtained
$E_{tres}$=14.44+3$\cdot$0.19=15.01~MeV. In the $^{16}$O nucleus, a
state indeed exists at the energy of $E_{\gamma}$=15.097~MeV with
\(J^{\pi} = 0^{+}\) \cite{15}. The discrepancy between the reaction
threshold estimate calculated from the two \(D\)-clusters of the
\(^{12}\text{C}\) nucleus and the experimental data can be explained
by the fact that the third cluster of the \(^{16}\text{O}\) nucleus
is less tightly bound, and therefore more massive, which shifts the
reaction threshold to an even higher photon energy. This estimate is
consistent with experimental data on the total cross-section for the
$^{16}$O photodisintegration into 4$\alpha$-particles \cite{16}. In
Ref.~\cite{16}, it was also concluded that the $^{16}$O nucleus does
not disintegrate into two $^8$Be nuclei. This can be explained by
the fact that, according to the Pauli exclusion principle, two {\it
S}-clusters cannot coexist within the nucleus.

Thus, tensor forces can shift the reaction threshold during cluster
emission. In addition to the examples of reactions with
$\alpha$-particle emission discussed above, similar effects may also
be observed in the emission of other composite particles.

\section{Conclusions}

In this work, it has been demonstrated that a realistic description
of nuclear properties for \(A > 4\) systems can be achieved without
introducing artificial fitting parameters, provided that the
intrinsic multi-dimensional structure of the nucleon-nucleon
interaction is strictly taken into account. By recognizing that
nuclear forces unfold within a four-dimensional generalized
configuration space governed by the coordinates (\(R\), \(L\),
\(S\), \(T\)), we provide a solid physical foundation for several
phenomena that conventional central-potential models fail to explain
adequately.

This approach provides an explanation for the $^8$Be lifetime, the
Hoyle state, the sequential reaction mechanism, and the reaction
threshold shift during cluster emission. Similar effects are
possible in other nuclei as well as in various nuclear reactions.

Thus, moving away from simplified spherically averaged potentials
toward a realistic, non-central tensor interaction operating across
multiple internal dimensions opens a promising alternative pathway
for nuclear structure theory.

The author expresses his gratitude to Dr. S.N. Afanasiev and Dr.
I.O. Afanasieva for discussing this work.


\begin{thebibliography}{5}

\bibitem{1} R.~Machleidt, F.~Sammarruca, and Y.~Song, Phys. Rev. C \textbf{53}, R1483 (1996).
 R. Machleidt, arXiv:nucl-th/0006014v1 (2000).

\bibitem{2} J.~Seely, A.~Daniel, D.~Gaskell, J.~Arrington, N.~Fomin et al.
Phys. Rev. Lett. \textbf{103}, 202301 (2009).

\bibitem{3} A.~Nogga, H.~Kamada, and W.~Glockle, Phys. Rev. Lett. \textbf{85},
944 (2000).

\bibitem{4} R.~B.~Wiringa, V.G.J.~Stoks, and R.~Schiavilla, Phys. Rev. C \textbf{51}, 38
(1995).

\bibitem{5} S.~A.~Coon, M.~D.~Scadron, P.~C.~McNamee, B.~R.~Barrett, D.W.E.~Blatt, and B.H.J.~McKellar, Nucl. Phys. \textbf{A317}, 242 (1979).

\bibitem{6} B.S.~Pudliner, V.R.~Pandharipande, J.~Carlson, Steven~C.~Pieper, and R.B.~Wiringa,
Phys. Rev. C \textbf{56}, 1720 (1997).

\bibitem{7} L.D.~Faddeev, Zh. Eksp. Teor. Fiz. [Sov. Phys. JETP \textbf{12}, 1041 (1961)].

\bibitem{8} O.A. Yakubovsky, Sov. J. Nucl. Phys. \textbf{5}, 937
(1967).

\bibitem{9} B.S.~Ishkhanov, M.E.~Stepanov, T.Y.~Tretyakova, Bulletin of Moscow Univ. \textbf{3}, 1, (in
Russian)(2014).

\bibitem{10} J.R.M.~Berriel-Aguayo and P.O.~Hess, Symmetry,
\textbf{12}(5), 738 (2020).

\bibitem{11} K.~Kravvaris, S.~Quaglioni, and P.~Navratil,
Phys. Rev. C \textbf{109}, 054603 (2024).

\bibitem{12} D.N.F.~Dunbar, R.E.~Picsley, W.A.~Wenzel, and W.~Whaling, Phys. Rev. \textbf{92},
3, (1953).

\bibitem{13} F.~Hoyle, Astrophys. J., Suppl. Ser. \textbf{1}, 121 (1954).

\bibitem{14} S.N.~Afanasiev, Ukr. J. Phys. \textbf{67}, 11 (2022).

\bibitem{15} F. Ajzenberg-Selove. Nucl. Phys. \textbf{A460}, 1,
(1986).

\bibitem{16} S.N.~Afanasiev. J. of Kharkiv Nat. Univ. \textbf{4}, \textnumero1025, 4 (in
Russian) (2012).


\end{thebibliography}
\end{document}